\documentclass{article}
\usepackage{amsmath}

\input{tcilatex}

\begin{document}

\begin{center}
\textbf{\large On non-equivalence of Lorentz and Coulomb gauges within
classical electrodynamics.}

Vladimir V. Onoochin

Sirius, 3A Nikoloyamski lane,  Moscow, 109004, Russia.\\[0pt]
e-Mail: {\underline {a33am@dol.ru}}\\[0pt]

\textbf{Abstract.}
\end{center}

It is shown that the well-known procedure for proving the equivalence of the
expressions for the electric field calculated using the Lorentz and Coulomb
gauges is incorrect. The difference between the two gauges is due to the
difference in the speed of propagation of a disturbance of the scalar
potential. As an auxiliary result, it is proven that the solution for the
electric field cannot be obtained directly from the Maxwell equations, i.e.
without introducing the scalar and vector potentials.

\begin{center}
\textbf{1. Introduction.}
\end{center}

\hspace{5mm} Recently, Tzontchev \textit{et al.} \cite{Ru} reported on an
experiment in which they detected a longitudinal component of the electric
field propagating at the speed of light in the near field of a radiator.
This result seems to be obvious because an electric field propagating with
the speed of light can easily be calculated, for example, by using Eq. 14.14
in \cite{JDJ}. However, the problem is that Eq. 14.14 was derived using the
Lorentz gauge in which disturbances of the scalar potential propagate with
the speed $c$. Because it follows from the experiment described in \cite{Ru}
that 0.95 of the total magnitude of the $\mathbf{E}$ field is created by the
scalar potential, it can be concluded that the scalar potential propagates
at the speed of light. However, this contradicts conventional electrodynamics.
It has been established in classical electrodynamics that the EM potential
cannot be treated as a physical quantity, but as a mathematical tool for
calculating EM fields \cite{JDJ} (Ch. 6.5, 3). Therefore, the solutions of
the wave equation can be chosen so that the speed of propagation $u$ of the
scalar potential can vary from zero to infinity. By choosing such solutions,
the gauge is also determined \cite{Chu}.

The most used gauges in electrodynamics are the Coulomb ( $u=\infty $) and
Lorentz ($u=c$) gauges. However, the infinite speed of propagation of the
scalar potential in the Coulomb gauge seems to contradict the theory of
relativity. Despite this, references [3, 2 (p 291, problem 6.20)] state that
the issues of causality and of the finite speed of propagation of
electromagnetic disturbances are obscured by the choice of the Coulomb
gauge: the potentials $\varphi (u)$ and $\mathbf{A}(u)$ are manifestly not
causal, but the fields can be shown to be. So it contradicts the conclusion
given in \cite{Ru} that ''\textit{The proper inference from this experiment
is that the Coulomb interaction cannot be considered as so called
'instantaneous action at a distance'}''( i.e. the scalar potential in the
Coulomb gauge). Actually, since we are only able to measure the EM fields
experimentally, we cannot draw any conclusions about the speed of
propagation of the scalar potential, if two solutions for the scalar and
vector potentials $\varphi (u_{1})$, $\mathbf{A}(u_{1})$ and $\varphi (u_{2})
$, $\mathbf{A}(u_{2})$, where $u_{i}$ is determined by the choice of gauge,
yield identical expressions for the electric field.

However, the results of the experiment described in \cite{Ru} suggest that
these expressions are not identical. So, one can question whether or not
different gauges in electrodynamics are actually equivalent. In this paper,
we consider this problem. Because existing studies of the equivalence of the
gauges have dealt with the Coulomb and Lorentz gauges, we will focus our
analysis on these gauges as well.

The plan of this paper is as follows. In section 2, we will calculate the
electric field in both gauges for the simplest model of the experimental
setup of \cite{Ru}. Also we will show that these calculations must be made
by means of the potentials but cannot be done by using the wave equation
for the $\mathbf{E}$ field derived from the Maxwell equations directly,
i.e. without introducing the potentials. Some explanation of the results
of the Sec. 2 is given in the sections 3 and 4 where  we will review the
derivation of the equivalence of the expressions for the electric field
calculated in both gauges and then we will show at what point this
derivation is wrong, i.e. the difference in the shapes of the elementary
classical charges calculated in both gauges is neglected. This difference
follows unambiguously from the expressions for the scalar potential in the
Coulomb and Lorentz gauges. Because its motion causes a change in the shape
of the charge, since the size of the charge contracts along its direction of
motion, one would expect that the greatest difference in the gauges should
occur in this direction as well. Finally, in section 5, we will draw some
conclusion about what physical effect is responsible for the difference in
the gauges.

\begin{center}
\textbf{2. An example of difference between the electric field calculated in
two gauges.}
\end{center}

\hspace{5mm} It is quite impossible to process complete calculations of the
E field detected by the antenna in the experiment of \cite{Ru}. So for
analysis of the fields in this system we should simplify the latter but in
such a way that its inherent features will be held. Therefore, we consider
the following simplification of the real experiment: a single charge moving
at the straight line which corresponds to the axis of symmetry of the
experimental installation. We assume too that this charge moves uniformly
which allows us to consider the most general case; i.e., when the properties
of the system do not depend on their initial conditions, and, therefore,
when choosing the advanced and retarded solutions as well.

Now we state the question: \textit{is it possible that the longitudinal
component of the E field calculated in different gauges has different
values?}. Similar calculations have not been made in \cite{Ru} so we wish
to make up this gap.

It should be noted that all formulas for calculations of the longitudinal
fields of the moving charge (in near non-radiative zone) are made for the
Lorentz gauge. So while calculating the electric field in the Coulomb gauge,
we use the method given in \cite{JRM}.

Thus, the equation for the vector potential in the Coulomb gauge is
(the Eq. 6.46 of \cite{JDJ})
\begin{equation*}
\nabla^{2} \mathbf{A}_{C} -\frac{\partial^{2}\mathbf{A}_{C}}{\partial t^{2}}
= - \frac{4\pi}{c} \mathbf{J}_{C} + \frac{1}{c}\nabla \frac{\partial
\varphi_{C}}{\partial t}
\end{equation*}

Because of independence of the current $\mathbf{J}_{C}$ and the scalar
potential $\varphi $ on each other, we are able to express the quantity
$\mathbf{A}_{C}$ in terms of the sum of two quantities; the first of which is
determined by one wave equation and the second from the other wave equation,
i.e.
\begin{equation}
\mathbf{A}_{C} = \mathbf{A}_{L} + \mathbf{A}_{\varphi }  \tag{2.1}
\end{equation}
\begin{equation}
\nabla^{2} \mathbf{A}_{L} -\frac{\partial^{2}\mathbf{A}_{L}} {c^{2}\partial
t^{2}} = - \frac{4\pi}{c} \mathbf{J}_{L}  \tag{2.2}
\end{equation}
\begin{equation}
\nabla^{2}\mathbf{A}_{\varphi } -\frac{\partial^{2} \mathbf{A}_{\varphi }}{%
c^{2}\partial t^{2}} = \frac{1}{c} \nabla \frac{\partial \varphi_{C}}{%
\partial t}  \tag{2.3}
\end{equation}

One can see from the Eq. 2.2 that $\mathbf{A}_{L}$ is the vector potential
in the Lorentz gauge. Now we find the difference between the electric fields
calculated in the Coulomb $\mathbf{E}_{C}$ and Lorentz $\mathbf{E}_{L}$
gauges
\begin{equation*}
\mathbf{E}_{C} - \mathbf{E}_{L} = \nabla [\varphi_{L}-\varphi_{C}] +\frac{%
\partial}{c\partial t}[\mathbf{A}_{L}-\mathbf{A}_{C}]=
\end{equation*}
\begin{equation}
= \nabla [\varphi_{L}-\varphi_{C}] - \frac{\partial \mathbf{A_{\varphi}}}{%
c\partial t} +\frac{\partial }{c\partial }\frac{1}{c} \int \frac{\mathbf{%
J}_{L}(r_{1})} {\left|\mathbf{r}_{1}-\mathbf{r} \right|}d\mathbf{r}_{1}-\frac{%
\partial }{c\partial }\frac{1}{c} \int \frac{\mathbf{J}_{C}(r_{1})} {%
\left|\mathbf{r}_{1} - \mathbf{r} \right|}d\mathbf{r}_{1}  \tag{2.4}
\end{equation}

Here we take into account that, as it will be shown in the Sec. 4, the
charges have different shapes in the Coulomb and Lorentz gauges so the
current densities $\mathbf{J}_{C}$ and $\mathbf{J}_{L}$ are different
as well. But this difference is asymptotically equal to zero (Appendix I),
so we omit third and fourth terms of \textit{rhs} of the above equation
from further consideration. 

We use Eq. 2.4 to calculate the difference between the electric fields in
the system defined above, i.e. \newline
the charge moves uniformly along the $\mathit{X}$-axis and the detector of
the electric field is on this axis as well.

$\mathit{X}$--------------------$q \to$
----------------Det-------------------------------

The first term on the \textit{rhs} of Eq. 2.4 is the difference between the
\textit{retarded} and \textit{instantaneous} scalar potentials. The magnitude
of the retarded potential of a uniformly moving point charge (if we measure
this quantity at the axis of motion of the charge) is
\begin{equation}
\varphi _{L}=1/\left| x-vt\right|   \tag{2.5}
\end{equation}

Eq. 2.5 is the reduced form of Eq. 21.39 in \cite{Fey} for ${\large {y=0}}$,
${\large {z=0}}$ , and the `current' time, but not in terms of the retarded
time. Eq. 2.5 coincides with the expression for the Coulomb potential of the
charge, when the charge is at the point ${\large {x-vt}}$, where ${\large {t}%
}$ is an instantaneous ('current') time. Therefore, the sum in the brackets
on the \textit{rhs} of Eq. 2.4 is equal to zero. For the electric fields
calculated in the Coulomb and Lorentz gauges to be equivalent, it is
necessary that the second term on the \textit{rhs} of the Eq. 2.4 be equal
to zero. But this is impossible if the terms, which are proportional to the
gradients of the scalar potentials, eliminate each other. One term on the
rhs of the Eq. 2.4 still remains and it can be expressed in terms of the
equation
\begin{equation}
\mathbf{E}_{C}-\mathbf{E}_{L}=\frac{\partial \mathbf{A}_{\varphi }}{%
c\partial t}  \tag{2.6}
\end{equation}
where $\mathbf{A}_{\varphi }$ is the solution of the wave equation with
source $\nabla \lbrack \partial \varphi _{C}/c\partial t\rbrack$.

To obtain solution of the Eq. 2.6, we will use the Lorentz procedure of
solving the wave equation (\cite{PPh} Ch 18.3); we do not refer to the
original work of Lorentz because he finds the solution for the fields and
not the potentials). In the Coulomb gauge, the distributed 'longitudinal
current' (the term $\nabla \lbrack \partial \varphi _{C}/c\partial t \rbrack$
instantaneously follows the charge creating this current, therefore the
Lorentz transformation of the coordinates reduces a static case and, as a
result, calculating the difference in the fields belonging to different
gauges reduces to solving a three dimensional integral.

Thus, the wave equation for $\mathbf{A_{\varphi }}$ is
\begin{equation}
\left( \frac{\partial ^{2}}{c^{2}\partial t^{2}}-\frac{\partial ^{2}}{%
\partial x^{2}}-\frac{\partial ^{2}}{\partial y^{2}}-\frac{\partial ^{2}}{%
\partial z^{2}}\right) \mathbf{A_{\varphi }}(x,y,z,t)=\frac{1}{c}\nabla
\frac{\partial \varphi _{C}(x-vt,y,z)}{\partial t}  \tag{2.7}
\end{equation}%
where we take into account that the \textit{rhs} of the Eq. 2.7 is formed
from the derivatives of scalar potential in the Coulomb gauge, where this
potential `instantaneously' follows the motion of the charge so that the $x$
and $t$ variables enter in the \textit{rhs} of the Eq. 2.7 in terms of the
combination $(x-vt)$.

Because the EM fields created by uniformly moving source must move with this
source too, the time and spatial derivatives are not independent of each
other, but are linked by the relation (the Eq. 18.10 of \cite{PPh})
\begin{equation*}
\frac{\partial }{\partial t}=-v\frac{\partial }{\partial x}
\end{equation*}

Therefore, the Eq 2.7 reduces to
\begin{equation}
\left( \left( 1-\frac{v^{2}}{c^{2}}\right) \frac{\partial ^{2}}{\partial
(x^{\prime })^{2}}+\frac{\partial ^{2}}{\partial y^{2}}+\frac{\partial ^{2}}{%
\partial z^{2}}\right) \mathbf{A_{\varphi }}(x^{\prime },y,z,t^{\prime })=-%
\frac{v}{c}\frac{\partial ^{2}\varphi _{C}(x^{\prime },y,z)}{\partial
(x^{\prime })^{2}}  \tag{2.8}
\end{equation}%
where $x^{\prime }=x-vt$. Since the \textit{rhs} of Eq. 2.8 does not depend
on time, the \textit{lhs} does not depend on time either, which means that
Eq. 2.8 reduces to the Poisson equation in elliptic coordinates. By changing
the variables $x^{\prime }/\sqrt{1-\frac{v^{2}}{c^{2}}}=\chi $ , Eq. 2.8
reduces to the ordinary Poisson equation with the \textit{rhs} containing a
spatially distributed source. Its solution is:
\begin{equation}
\mathbf{A_{\varphi ,X}}=\frac{v}{c(1-\frac{v^{2}}{c^{2}})}\int \frac{%
\partial ^{2}\varphi (\chi ,y,z)/\partial \chi ^{2}}{\left| \mathbf{R}_{1}
-\mathbf{r}(\chi ,y,z)\right| }d\chi dydz
\tag{2.9}
\end{equation}%
where
\begin{equation*}
R_{1}=\sqrt{\left( 1-v^{2}/c^{2}\right) \left( X-vt\right) ^{2}+Y^{2}+Z^{2}}
\end{equation*}

Inserting the expression for $\mathbf{A_{\varphi ,X}}$ (Eq. 2.9) into Eq.
2.6, we finally obtain
\begin{equation*}
\mathbf{E}_{C}(\mathbf{R},t)-\mathbf{E}_{L}(\mathbf{R},t)=
\frac{v}{c(1-\frac{v^{2}}{c^{2}})}\frac{\partial }
{c\partial t}\int \frac{\partial ^{2}\varphi (\chi,y,z)/\partial \chi ^{2}}
{\left| \mathbf{R}_{1}-\mathbf{r}(\chi ,y,z)\right| }d\chi dydz
\end{equation*}

One can easily see that because the integrand is not a symmetric expression,
the integral over the whole space is not equal to zero (we do not finish the
calculation of this integral because its concrete form is not essential).
Therefore, we find that the field is actually different in the different
gauges.

Here, one can expect an objection that because the $\mathbf{E}$ field can be
calculated directly from Maxwell equations, the analysis of the difference
in the $\mathbf{E}$ fields calculated in both gauges loses its sense.
However, it is not so. We show that for this case, i.e. the case of
\textit{longitudinal} fields, it is impossible to obtain the solution for
the $\mathbf{E}$ field without using the EM potentials.

To avoid any cumbersome calculations which can be caused by necessity to
describe radiation processes, we consider simplest electrodynamical system
which one is given above, i.e. the charge moves uniformly along the
$\mathit{X}$ axis. In description of this system, we will be able to obtain
the expressions for the field in explicit form which allows to compare
the solutions for the $\mathbf{E}$ field obtained in two ways.

Firstly, we consider derivation of direct, i.e. made without introducing the
potentials, wave equation (DWE) for $\mathbf{E}$ field. Using two Maxwell
equations (second and fourth Eqs. 6.28 of \cite{JDJ})
\begin{equation}
\nabla \times \mathbf{E} = - \frac{\partial \mathbf{H}}{c\partial t}
\tag{2.10}
\end{equation}
\begin{equation}
\nabla \times \mathbf{H} =+\frac{\partial \mathbf{E}}{c\partial t}
+\frac{4\pi \mathbf{J}}{c}
\tag{2.11}
\end{equation}

Taking the curl of Eq. 2.10 and partial time derivative, divided by $c$,
of Eq. 2.11, we obtain
\begin{equation*}
\nabla\times\nabla\times \mathbf{E}
= - \frac{\partial \nabla \times \mathbf{H}}{c\partial t}
\end{equation*}
\begin{equation*}
\frac{\partial \nabla \times \mathbf{H}}{c\partial t} =
+ \frac{\partial ^2\mathbf{E}}{c^2\partial t^2}
+ \frac{4\pi \partial \mathbf{J}}{c^2\partial t}
\end{equation*}

Eliminating the $\mathbf{H}$ field from the above equations, we have
\begin{equation}
\nabla \times \nabla \times \mathbf{E} + \frac{\partial ^2\mathbf{E}}
{c^2\partial t^2} = - \frac{4\pi \partial \mathbf{J}}{c^2\partial t}
\tag{2.12}
\end{equation}

Substituting the vector identity
\begin{equation*}
\nabla \times \nabla \times \mathbf{E} =
\nabla \left( \nabla \cdot \mathbf{E} \right) - \nabla ^2\mathbf{E}
\end{equation*}
to the Eq. 2.12, we obtain
\begin{equation*}
- \nabla ^2\mathbf{E} +\frac{\partial ^2\mathbf{E}}{c^2\partial t^2}
= - \frac{4\pi \partial \mathbf{J}}{c^2\partial t} -
\nabla \left( \nabla \cdot \mathbf{E} \right)
\end{equation*}

From the first of Eqs. 6.28 of \cite{JDJ},
$\nabla \cdot \mathbf{E} = 4\pi \rho $, which gives
\begin{equation}
 - \nabla ^2\mathbf{E} + \frac{\partial ^2\mathbf{E}}{c^2\partial t^2}
= - \frac{4\pi \partial \mathbf{J}}{c^2\partial t} - 4\pi \nabla \rho
\tag{2.13}
\end{equation}

Now we use the Eq. 14 for calculation of the electric fields created by
the elementary charge uniformly moving along the $\mathit{X}$ axis.

Since the wave operator $ - \nabla ^2 + \left( \partial ^{2}...
/c^2\partial t^2 \right)$
is a scalar, direction of the $\mathbf{E}$ vector is defined by direction of the vector
of the source, i.e. of $-\left( 4\pi \partial \mathbf{J}/c^2\partial t
\right) - 4\pi \nabla \rho $. Now we use the principle of superposition
and present the source as four separate sources directed along the axes
($x$, $y$, $z$).
\begin{equation*}
\left(  - \frac{4\pi }{c^2}\frac{\partial J_{x} }{\partial t};-4\pi
\frac{\partial \rho }{\partial x};-4\pi \frac{\partial \rho }{\partial
y};-4\pi \frac{\partial \rho }{\partial z} \right)
\end{equation*}
The total $\mathbf{E}$ field can be presented as a sum of four independent
fields, each of them is a solution of the wave equation
\begin{equation}
 - \nabla ^2E_{x,J} + \frac{\partial ^2E_{x,J} }{c^2\partial t^2} = -
\frac{4\pi \partial J_{x} }{c^2\partial t}
\tag{2.14a}
\end{equation}
\begin{equation}
 - \nabla ^2E_{x,\rho } + \frac{\partial ^2E_{x,\rho } }{c^2\partial t^2} =
-4\pi \frac{\partial \rho }{\partial x}
\tag{2.14b}
\end{equation}
\begin{equation}
 - \nabla ^2E_{y,\rho } + \frac{\partial ^2E_{y,\rho } }{c^2\partial t^2} =
-4\pi \frac{\partial \rho }{\partial y}
\tag{2.14c}
\end{equation}
\begin{equation}
 - \nabla ^2E_{z,\rho } + \frac{\partial ^2E_{z,\rho } }{c^2\partial t^2} =
-4\pi \frac{\partial \rho }{\partial z}
\tag{2.14d}
\end{equation}
To obtain the solution of the Eq. 2.14b, we use the Green formula
(the Eq. 6.66 of \cite{JDJ} with the `\textit{source}'
\begin{equation*}
f(\mathbf{r}',t')= \left( \partial \rho /\partial x'\right)
\end{equation*}
i.e.
\begin{equation}
E_{x,\rho } (\mathbf{r},t) = \int \frac{\left( \partial \rho /
\partial x' \right)_{ret} }{\left| \mathbf{r}-\mathbf{r}' \right|}d\mathbf{r}'
\tag{2.15}
\end{equation}
where $r$ is the radius vector of the point of detection of
the fields and note `\textit{ret}' means that the function
$\left( \partial \rho / \partial x' \right)$
should be calculated at retarded time.

To calculate the integral 2.15 in the limit of point charge, we should make
integration by parts of the \textit{rhs} of the above equation.
\begin{equation}
E_{x,\rho } \left( \mathbf{r},t \right) =
\int \frac{\left( \partial \rho /\partial x' \right)_{ret}}
{\left| \mathbf{r} - \mathbf{r}' \right|} d\mathbf{r}'
= - \int \rho _{ret} \frac{\partial }{\partial
x'}\frac{1}{\left| \mathbf{r} - \mathbf{r}' \right|} d\mathbf{r}'
= \int \rho _{ret} \frac {x - x'}{\left| \mathbf{r}
- \mathbf{r}' \right|^{3}} d\mathbf{r}'
\tag{2.16}
\end{equation}
Now, while calculating the integral 2.16, we should take into account that
the charge is a non-point object; so after going in integration over
the volume occupated by the elementary charge to integration over the
charge itself, we have (all details of transition from $dr'$ integration
to $de$ integration are given in Ch. 18.1 of \cite{PPh})
\begin{equation}
E_{x,\rho } \left( \mathbf{r},t \right) = \int \frac {\left( x - x'\right)}
{\left| \mathbf{r} - \mathbf{r}'\right|^{3}}\rho _{ret}d\mathbf{r}' =
\int \frac {\left( x - x'\right)}{\left| \mathbf{r} - \mathbf{r}'\right|^{3}}
\frac{de}{\left[ r-(\mathbf{r} \cdot \mathbf{v})/(cr) \right]}
\tag{2.17}
\end{equation}

As a result, we obtain
\begin{equation*}
E_{x,\rho } (\mathbf{r},t) = \frac{qx}{r^3\left[ 1 - \left( \mathbf{r}
\cdot \mathbf{v} \right) / (cr) \right]}_{ret}
\end{equation*}

We will not calulate transversal terms because to show incorrectness of the
DWE solutions, it is sufficient to obtain for the only component of the
$\mathbf{E}$ field that this component obtained from the DWE and from
the LW potential is different.

However, the electric field is created not only by the charge but by the
current density too. So we take into account the solution of Eq. 2.14a.
One can see that the \textit{rhs} of Eq. 2.14a
$ -\left(4\pi / c^2 \right) \left( \partial j_x / \partial t \right)$
can be changed by the term
$ - \left(4\pi v^2/c^2 \right) \left( \partial \rho /\partial x \right)$
in case of uniformly moving charge. So the total $E_{x}$ solution of the
DWE is 
\begin{equation}
E_x (\mathbf{r},t)=\left({1 - \frac{v^2}{c^2}} \right)
\frac{qx}{r^3\left( {1 - \left( \mathbf{r} \cdot \mathbf{v}) \right)
 /(cr)} \right)}_{ret} \tag{2.18}
\end{equation}
and similar field calculated after the LW potential (in longitudinal
direction there is no radiated term) is
\begin{equation}
E_x (\mathbf{r},t) = -\left( {1 - \frac{v^2}{c^2}} \right)
\frac{\partial }{\partial x}\frac{q}{\left[ {r - \left( \mathbf{r} \cdot
\mathbf{v} \right) / c} \right]}_{ret}
\tag{2.19}
\end{equation}
Obviously, Eqs. 2.18 and 2.19 are \textit{different}. So if we assume that
the Eq. 2.18 is correct we must assume that the Eq. 2.19 is incorrect. However,
the Eq. 2.19 is a part of general formula for the E fields of arbitrary moving
charge. So our assumption will require radical revising the basic formulas
of the classical electrodynamics, therefore, we must conclude that the DWE
for the electric field gives incorrect result.

But this \textit{strange result} of difference in the $\mathbf{E}$
fields calculated via the potentials but in different gauges must be explaned.
So one can suggest that there is a some error in the proof of equivalence of
the electrodynamical gauges. Below we will show that this suggestion has some
ground but before we review existing proof of this equivalence within the
classical electrodynamics.

\begin{center}
\textbf{3. Derivation of the expressions for the $E$ field calculated in
both gauges.}
\end{center}

\hspace{5mm} One can assume that a sufficient condition of equivalence for
both gauges is the identical form of the expressions for the electric field
calculated by both these gauges. Proof of this is well represented in the
scientific literature (see \cite{BG}, \cite{JDJ} 2$^{nd}$ and 3$^{rd}$
editions, and \cite{JRM}). In spite of this, we recall that these derivations miss a critical
point. Although our derivation does not coincide completely to those given
in \cite{BG}, \cite{JDJ},  \cite{JRM}, we keep the basic ideas used in the
cited works.

Thus, we consider the wave equations for the vector and scalar potentials in
the Coulomb and Lorentz gauges, respectively. The wave equations for the
vector potential in the Coulomb (for $\mathbf{A}_{C}$) and Lorentz (for 
$\mathbf{A}_{L}$) gauges are (Eqs. 3 and 6 of \cite{JRM}, where we use the
same notation used in \cite{JRM}):
\begin{equation}
\nabla ^{2}\mathbf{A}_{C}-\frac{\partial ^{2}\mathbf{A}_{C}}{c^{2}\partial
t^{2}}=-\frac{4\pi }{c}\mathbf{J}+\frac{1}{c}\nabla \frac{\partial \varphi
_{C}}{\partial t}  \tag{3.1}
\end{equation}%
\begin{equation}
\nabla ^{2}\mathbf{A}_{L}-\frac{\partial ^{2}\mathbf{A}_{L}}{c^{2}\partial
t^{2}}=-\frac{4\pi }{c}\mathbf{J}  \tag{3.2}
\end{equation}

Subtracting Eq. 3.2 from Eq. 3.1, we obtain (Eq. 11 of \cite{JRM}):
\begin{equation}
\nabla ^{2}[\mathbf{A}_{C}-\mathbf{A}_{L}]-\frac{1}{c^{2}}\frac{\partial
^{2}[\mathbf{A}_{C}-\mathbf{A}_{L}]}{\partial t^{2}}=\frac{1}{c}\nabla \frac{%
\partial \varphi _{C}}{\partial t}  \tag{3.3}
\end{equation}

The corresponding equations for the scalar potential in the Coulomb (for $%
\varphi _{C}$) and Lorentz (for $\varphi _{L}$) gauges are (Eqs. 4 and 7 of
\cite{JRM}):
\begin{equation}
\nabla ^{2}\varphi _{C}=-4\pi \rho   \tag{3.4}
\end{equation}%
\begin{equation}
\nabla ^{2}\varphi _{L}-\frac{1}{c^{2}}\frac{\partial ^{2}}{\partial t^{2}}%
\varphi _{L}=-4\pi \rho   \tag{3.5}
\end{equation}

Using the above two equations, we take their difference and find that the
term on the rhs of the Eqs. 3.4 and 3.5, corresponding to the charge
density, is eliminated. However, another term appears, which corresponds to
the second time derivative of $\varphi _{L}$:
\begin{equation}
\nabla ^{2}[\varphi _{C}-\varphi _{L}]-\frac{1}{c^{2}}\frac{\partial
^{2}[\varphi _{C}-\varphi _{L}]}{\partial t^{2}}=-\frac{1}{c^{2}}\frac{%
\partial ^{2}\varphi _{C}}{\partial t^{2}}  \tag{3.6}
\end{equation}

Now we transform Eqs. 3.3 and 3.6 in such a way that their \textit{rhs}'
will have the identical form. To do it, we apply the gradient operator to
Eq. 3.6 and operator $[\partial ../c\partial t]$ to Eq. 3.3. As a
result, we find, after commuting the gradient and the operator $[\partial ../
c\partial t]$ with the wave operator, that
\begin{equation}
\nabla ^{2}[\nabla (\varphi _{L}-\varphi _{C})]-\frac{1}{c^{2}}\frac{%
\partial ^{2}[\nabla (\varphi _{L}-\varphi _{C})]}{\partial t^{2}}=\frac{1}{%
c^{2}}\nabla \frac{\partial ^{2}\varphi _{C}}{\partial t^{2}}  \tag{3.7}
\end{equation}%
\begin{equation}
\nabla ^{2}[\partial (\mathbf{A}_{C}-\mathbf{A}_{L})/c\partial t]-\frac{1}{%
c^{2}}\frac{\partial ^{2}[\partial (\mathbf{A}_{C}-\mathbf{A}_{L})/c\partial
t]}{\partial t^{2}}=\frac{1}{c^{2}}\nabla \frac{\partial ^{2}\varphi _{C}}{%
\partial t^{2}}  \tag{3.8}
\end{equation}%
which are similar to Eqs. 24 and 25 of \cite{JRM}. From Equations (3.7) and
(3.8), both $\nabla (\varphi _{L}-\varphi _{C})$ and
$\partial (\mathbf{A}_{C}-\mathbf{A}_{L})/c\partial t$ satisfy the same
differential equation. Therefore,
\begin{equation}
\nabla (\varphi _{L}-\varphi _{C})=\frac{\partial (\mathbf{A}_{C}-\mathbf{A}%
_{L})}{c\partial t}  \tag{3.9}
\end{equation}

Transforming Eq. 3.9 and using the definition for the electric field, we
have
\begin{equation}
\mathbf{E}_{C}=-\nabla \varphi _{C}-\frac{\partial \mathbf{A}_{C}}{c\partial
t}=-\nabla \varphi _{L}-\frac{\partial \mathbf{A}_{L}}{c\partial t}=\mathbf{E%
}_{L}  \tag{3.10}
\end{equation}%
i.e. the equivalence of the expressions for the $\mathbf{E}$ field in both
gauges is proven.

We note that Eq. 3.10 is a constructive method for calculatimg the $\mathbf{E%
}$ field in the Coulomb gauge \cite{JRM}: scalar Coulomb potential, entering
in Eq. 3.10, is calculated as a solution of the Poisson equation and that
part of the $\mathbf{E}$ field created by the vector potential is determined
by using the following form of Eq. 3.10
\begin{equation}
\frac{\partial \mathbf{A}_{C}}{c\partial t}=\nabla \varphi _{L}-\nabla
\varphi _{C}+\frac{\partial \mathbf{A}_{L}}{c\partial t}  \tag{3.11}
\end{equation}

Thus, the total electric field is presented as a superposition of rotational
and irrotational components which is important for analysis of the fields
near the radiator \cite{JRM}. Equivalence of the magnetic field in both
gauges follow from the equation
\begin{equation*}
\nabla \times \mathbf{A}_{C}=\nabla \times \mathbf{A}_{L}
\end{equation*}%
since $\mathbf{A}_{L}$ differs from $\mathbf{A}_{C}$ \ by only the gradient
of some scalar function.

\begin{center}
\textbf{4. Analysis of the derivation presented in the Sec. 3.}
\end{center}

\hspace{5mm} It is necessary to say that despite the obviousness of the
proof presented above, it contains a few mistakes. First, to prove
equivalence in a mathematically strict way, one must analyze the expressions
for the electric fields and not the differential equations for these fields.
If one focuses on analysis of the latter, one must take into account the
initial and boundary conditions, because solutions of identical equations,
but for different boundary and initial conditions, are different. This point
is missed in the existing proof.

The second missing point is in the procedure of proof itself, i.e. when
\textit{rhs} of equations 3.1, 3.2, 3.4 and 3.5 is eliminated, one does not
consider that the functions describing the current and charge densities in
the Coulomb and Lorentz gauges are \textit{different}. This fact can be
established by using the idea of Lorentz to find that the sizes of uniformly
moving charge, which contract along their direction of motion (this
procedure developed by Lorentz is described with more clarity in \cite{Fr}).
Lorentz found that the equipotential surfaces of the scalar
Liennard-Wiechert potential, expressed in terms of the coordinates of the
frame of reference where it is assumed that the observer is at rest and the
charge is moving, i.e. $\varphi (r,t)=const$, are ellipsoids of rotation
contract along the axis of motion of the charge. Since, as Lorentz
concluded, the surface of the charge is defined to be an equipotential
surface, this surface must have an ellipsoidal shape in this frame too.

Following Lorentz's procedure, we will show that the functions $\rho $ and $%
\mathbf{J}$, describing the charge and current densities of the elementary
charge, are different in these gauges. But first, we must define what an
elementary charge in the classical electrodynamics is.

It is a widespread opinion in classical electrodynamics, that we are able to
assume point charges only. At least, any calculations of electrodynamical
quantities cannot be based on a specific distribution of the charge inside
the electron. However, from a physical point of view, it is impossible to
treat the classical electron as a point particle because it leads to
divergences in the theory (runaway solutions, etc., see, for example, 
\cite{Rohr}). Therefore, according to the recommendations given in \cite{PPh}
(beginning of Ch. 18.1), we assume that the radius of classical charge is
\textit{finite} and we associate a physical meaning to those properties of
the electrodynamical system which do not depend on the radius of the charge.

Thus, we have \underline{Statement I}:\newline
\textit{A surface of the elementary charge is the surface for which the
condition
\begin{equation*}
\varphi (x_{0},y_{0},z_{0})=const \tag{4.1}
\end{equation*}%
is fulfilled}\newline
(where, $x_{0},y_{0},z_{0}$ are the coordinates of the surface of the
charge). So, this surface is an equipotential surface. We note that for a
moving charge, the lines of $\mathbf{E}$ field are not normal to the
equipotential surface ( Eq. 4.1), since we must take into account not only
the term
\begin{equation*}
\mathbf{E}=-\nabla \varphi
\end{equation*}%
but rather the entire expression
\begin{equation}
\mathbf{E}=-\nabla \varphi -\frac{\partial \mathbf{A}}{c\partial t}
\tag{4.2}
\end{equation}
Due to the last term in Eq. 4.2, the $\mathbf{E}$ field lines are no longer
normal to the $\varphi $ surfaces.

For the Lorentz gauge, it is proven in \cite{PPh}, and for the Coulomb gauge,
we prove it in the Appendix II. We emphasize that the results presented
Appendix II, i.e. the spherical shape of a moving charge in the Coulomb
gauge changes, which contradicts to relativistic theory because according to
the latter, \textit{any} charge should contract. But our 'strange result'
is caused by using the Coulomb gauge which is essentially non-relativistic
so some quantities calculated in this gauge have no relativistic properties.

It is necessary to point out that the `relativistic suggestion' that a
moving charge in the Coulomb gauge must contract also cannot be checked
experimentally. It results from the following: \newline
1. the shape of the surface of the charge can be determined by measureing
the fields or by calculating the potentials since direct measurement (not
via field quantities) is \textit{impossible} (there is no 'charge-charge'
interaction);\newline
2. because the lines of $\mathbf{E}$ field are not normal to the surface of
the moving charge, one cannot use direct measurement of the EM fields to
reconstruct the shape of the surface.

So the only way to obtain information about the shape of this surface is to
do it via calculation of the $\varphi $ potential, as it has been made by
Lorentz, and we will using his method.

Now we have \underline{Statement II}\newline
\textit{for both gauges, the equipotential surfaces, i.e. those ones meeting
the condition
\begin{equation*}
\varphi (r,t)=const
\end{equation*}%
for different $r$ and given instant $t$ are the concentric surfaces
converging to the limiting point, which is the center of the elementary
charge}.\newline
For the Coulomb gauge this Statement follows from rotational symmetry of the
expression for the scalar potential (Eq. 6.45 of \cite{JDJ}) and for the
Lorentz gauge, it follows from the Eq. 18.20, but in the latter case, we
must calculate the shapes of the surfaces separately when the point of
observation is outside the charge and when this point is inside the
elementary charge.

It follows from the Statement II that\newline
\underline{Consequence I}:\newline
\textit{While $r\rightarrow 0$ the set $\varphi (r,t)=const$ forms a
geometric sequence} (the sequence of converging surfaces).\newline
\underline{Consequence II}:\newline
\textit{For different gauges, these surfaces are different and for any $r$}
\begin{equation}
\varphi _{C}(r,t)=C_{1}  \tag{4.3}
\end{equation}%
\begin{equation}
\varphi _{L}(r,t)=C_{2}  \tag{4.4}
\end{equation}%
where $C_{1}$ and $C_{2}$ are constants; \textit{and for any $C_{1}$ and $%
C_{2}$}
\begin{equation}
\varphi _{C}(r,t)\neq \varphi _{L}(r,t)  \tag{4.5}
\end{equation}%
The Eq. 4.5 can be easily proven. Because $\varphi _{C}(r,t)$ and $\varphi
_{L}(r,t)$ are solutions of \textit{different equations} (Eqs. 3.4 and 3.5),
they must be different too. Strictly speaking, the intersection of the two
surfaces the Eqs. 4.3 and 4.4, yields some curve but the coincidence of
these surfaces is never possible.\newline
\underline{Consequence III}:\newline
\textit{For all gauges, the limiting point of converging sequences is
unique, it is the point of center of the elementary charge}

Now we choose the parameter $R_{0}$ as a 'radius' of moving elementary
charge. In the Coulomb gauge, $R_{C,0}=\sqrt{%
(x_{0})^{2}+(y_{0})^{2}+(z_{0})^{2}}$, and in the Lorentz gauge $R_{L,0}=%
\sqrt{(x_{0})^{2}/\left( 1-v^{2}/c^{2}\right) +(y_{0})^{2}+(z_{0})^{2}},$
where $x_{0}...$ are defined above. Actually, $\varphi _{L}$ depends on the
coordinates $x_{0}...$ not via $R_{L,0}$ but\ rahter via some other
combination of the these variables, but since this specific dependence is
not important for our procedure, we schematically write this dependence via $%
R_{L,0}$. We do not know the exact values for $R_{C,0}$ and $R_{L,0}$, we
only know that both $R_{0}\rightarrow 0$ but both $R_{0}\neq 0,$ which
corresponds to the definition for the radius of the elementary charge given
in \cite{PPh}.

It follows from Statement I that the shape of the charge in the Coulomb
gauge is described by
\begin{equation}
\varphi _{C}(R_{C,0},t)=C_{3}  \tag{4.6}
\end{equation}%
and the shape of the charge in the Lorentz gauge is described by
\begin{equation}
\varphi _{L}(R_{L,0},t)=C_{4}  \tag{4.7}
\end{equation}%
where $C_{3}$ and $C_{4}$ are some constants; i.e. Eq. 4.6 belongs to the
sequence in Eq. 4.3 and Eq. 4.7 to the sequence in Eq. 4.4, respectively.
However, due to the inequality in Eq. 4.5, the equation
\begin{equation*}
\varphi _{C}(R_{C,0},t)=\varphi _{L}(R_{L,0},t)
\end{equation*}%
cannot be fulfilled \textit{for any $R_{C,0}$ and $R_{L,0}$}. Physically it
means that the shapes of the charge in different gauges are different too
and, therefore, $\rho _{L}(r)$ and $\rho _{C}(r)$ are not identical and the
mathematical operation of subtracting one function from the other yields a
non-zero result. Because the above proof does not depend on specific values
of $R_{0},$ it is correct in the limit of a point charge too.

A further consideration is trivial. Taking into account that the functions $%
\mathbf{J}$ and $\rho $  are different in different gauges, we obtain
\begin{equation}
\nabla^{2} \mathbf{A_{C}} -\frac{\partial^{2} \mathbf{A}_{C}} {c^{2}\partial
t^{2}} = - \frac{4\pi}{c} \mathbf{J}_{C} + \frac{1}{c}\nabla \frac{\partial
\varphi_{C}}{\partial t}  \tag{4.8}
\end{equation}
\begin{equation}
\nabla^{2} \mathbf{A}_{L} -\frac{\partial^{2} \mathbf{A}_{L}} {c^{2}\partial
t^{2}} = - \frac{4\pi}{c} \mathbf{J}_{L}  \tag{4.9}
\end{equation}

The analogue of Eq 3.3 is
\begin{equation}
\nabla^{2} [\mathbf{A}_{C} - \mathbf{A}_{L}] -\frac{1}{c^{2}} \frac{%
\partial^{2} [\mathbf{A}_{C} - \mathbf{A}_{L}]}{\partial t^{2}} = \frac{1}{c}%
\nabla \frac{\partial \varphi_{C}}{\partial t} + \frac{4\pi }{c}[\mathbf{J}%
_{L} - \mathbf{J}_{C}]  \tag{4.10}
\end{equation}

Applying the wave equation for the scalar potential in a similar way
\begin{equation}
\nabla^{2} \varphi_{C} = - 4\pi \rho_{C}  \tag{4.11}
\end{equation}
\begin{equation}
\nabla^{2} \varphi_{L} -\frac{1}{c^{2}} \frac{\partial^{2}}{\partial t^{2}}
\varphi_{L} = - 4\pi \rho_{L}  \tag{4.12}
\end{equation}
we obtain
\begin{equation}
\nabla^{2} [\varphi_{C} - \varphi_{L}] -\frac{1}{c^{2}} \frac{\partial^{2}
[\varphi_{C} - \varphi_{L}]}{\partial t^{2}} = -\frac{1}{c^{2}}\frac{%
\partial^{2} \varphi_{C}}{\partial t^{2}} - 4\pi [\rho_{C} - \rho_{L}]
\tag{4.13}
\end{equation}

Thus, we find that the wave equations for the quantities $\nabla (\varphi
_{L}-\varphi _{C})$ and $\partial (\mathbf{A}_{C}-\mathbf{A}_{L})/c\partial t$
coincide, provided the condition
\begin{equation}
\frac{\partial (\mathbf{J}_{C}-\mathbf{J}_{L})}{c^{2}\partial t}-\nabla
(\rho _{C}-\rho _{L})=0  \tag{4.14}
\end{equation}%
is satisfied. But in general, this is not the case and for uniformly moving
charge, the \textit{lhs} of the Eq 4.14 reduces to
\begin{equation}
(1-v^{2}/c^{2})\nabla (\rho _{C}-\rho _{L})  \tag{4.15}
\end{equation}

It seems, however, that there is one more way to prove the equivalence of
the expressions for the electric field, because the non-zero term on the
\textit{rhs} of the wave equation is not equal to zero only in the area
occupied by the charge itself. So we can expect that, after integration of
the wave equation, the non-compensated term in Eq. 4.15 will tend to zero,
while receding the point of observation from the charge.

It is expressed in explicit form as (for simplicity we consider the case of
a uniformly moving charge, where all details of the calculations are given
in Appendix I):
\begin{equation}
\mathbf{E}_{C}(R)-\mathbf{E}_{L}(R)=4\pi (1-v^{2}/c^{2})\int \frac{\nabla
_{r}[\rho _{C}(r)-\rho _{L}(r)]}{\left| \mathbf{R}-\mathbf{r}\right| }
d\mathbf{r}
\tag{4.16}
\end{equation}
where the integral is calculated for the retarded time. Since the integral
of the charge density over the whole space is equal to the total charge in
both gauges, it is easy to show that the above expression rapidly tends to
zero when $R>>a$, where $a$ is the radius of the elementary classical charge
(Appendix I), i.e. the expressions for the electric field are
asymptotically equivalent in both gauges, and if one takes into account the
radius of the classical charge, the limiting area of integration in the Eq.
4.16 should be set to zero, the gauges are equivalent in classical
electrodynamics.

But one must take into account that the term $-[\partial ^{2}\varphi
_{C}/c^{2}\partial t^{2}]$ is artificially added to the \textit{rhs} and
\textit{lhs} of the wave equation ( Eq. 3.4). Since this term is added to
the both the right and the left sides of the equations, it is mathematically
correct. But if this term is used on the \textit{lhs} of the equation to
construct the Green's function and the same term in the rhs of the equation
is used as a source for the Green's function, which means that the \textit{%
lhs} and \textit{rhs} of the equation are being treated differently, it is
absolutely incorrect. As a consequence of this incorrect procedure, it leads
to a difference in the expressions for the calculated fields.

\begin{center}
\textbf{5. Conclusion.}
\end{center}

\hspace{5mm} It would be interesting to do an analysis as to why such a
microscopic effect as the changing the shape of electric charge (for a
uniformly moving electron, its surface becomes elliptical in the Lorentz
gauge and remains spherical in the Coulomb gauge) causes a macroscopic
effect (difference in the fields). It is especially strange since formally
we are able to decrease the radius of the charge to zero. So from our point
of view, the macroscopic effect is \ not caused by the changing shape of the
elementary classical charge, but rather by properties of the aether:
finiteness (of infiniteness) of the speed of the scalar EM interaction
determines the magnitude of both the EM fields and the shape of the charge
creating these fields.

Thus, just the act of defining the speed of propagation of the scalar EM
interaction in a medium (aether or vacuum) defines the correct gauge for
this system, as well as the shape of the elementary classical charge.
Because we have an example of the reverse influence of the medium on the
charge (in \ the case of uniform motion of the charge we have some
equilibrium process for converging and diverging EM waves), this influence
unambiguously determines the equilibrium shape of the moving charge. The
mechanism by which the medium influences the charge is still unexplained,
but within the framework of this effort, it is impossible to find an
explanation. It should be noted that in relativistic theory, the term
'medium' is not used, so we use the term `aether' but we do not make any
claims about its reality.

It would be noted that one of the aims of this work is to turn the
scientific community's attention to the fact that until now, some problems
of electrodynamics, which seemed to be absolutely irrefutable, cannot be
conclusively solved. So it would be interesting to re-examine some of the
ideas of Whittaker(\cite{Wh}, also see \cite{Dv}), especially regarding the
formation of the Coulomb (or scalar) potential from convergent and divergent
EM waves. Finally, the difference in the properties of the scalar potential
calculated in the Coulomb and Lorentz gauges gives differences for all
other field quantities.

Thus, the final conclusion of this work is that \textit{in classical
electrodynamics}, the uniqueness of the description of some systems requires
setting not only the initial and boundary conditions but also \textit{the
speed of propagation} of the scalar potential as well, where the latter
unambiguously determines the gauge which we must use while obtaining
solutions for the EM fields.

\begin{center}
\textbf{Acknowledgments}
\end{center}

The authors are indebted to Prof. Joseph R. Mautz for discussion and
critical comments. Also the auhtors wish to express their gratitudes to Drs.
Andrew E. Chubykalo and Rumen I. Tzontchev for detailed discussion
concerning the results of work \cite{Ru}.

\begin{center}
\textbf{Appendix I. Derivation of the Eq. 4.16}
\end{center}

Taking the gradient of both sides of the Eq. 4.13 and partial time derivative
(divided on $c$) of both sides of the Eq. 4.10, we obtain
\begin{equation}
\nabla^{2} [\nabla (\varphi_{C} - \varphi_{L})] -\frac{1}{c^{2}} \frac{%
\partial^{2}[\nabla (\varphi_{C}-\varphi_{L})]}{\partial t^{2}} =-\frac{1}{%
c^{2}}\nabla \frac{\partial^{2} \varphi_{C}}{\partial t^{2}} + \nabla 4\pi[%
\rho_{C}-\rho_{L}]  \tag{AI.1}
\end{equation}
\begin{equation}
\nabla^{2} [\partial(\mathbf{A_{C}} - \mathbf{A_{L}})/\partial t] -\frac{1}{%
c^{2}} \frac{\partial^{2}[\partial(\mathbf{A_{C}} - \mathbf{A_{L}})/\partial
t]} {\partial t^{2}} = \frac{1}{c^{2}}\nabla \frac{\partial^{2} \varphi_{C}}{%
\partial t^{2}} -\frac{\partial }{c\partial t}\frac{4\pi}{c} [\mathbf{J_{C}}
- \mathbf{J_{L}}]  \tag{AI.2}
\end{equation}

Now we are able to form, using the sum of the Eqs. AI.1 and AI.2, the wave
equation for difference between the electric field $\mathbf{E_{C}}$ and $%
\mathbf{E_{L}}$:
\begin{equation}
\nabla^{2} [\mathbf{E_{C}} - \mathbf{E_{L}}]-\frac{1}{c^{2}} \frac{%
\partial^{2}[\mathbf{E_{C}} - \mathbf{E_{L}}]} {\partial t^{2}} = \nabla 4\pi%
[\rho_{C}-\rho_{L}] -\frac{\partial }{c\partial t}\frac{4\pi}{c} [\mathbf{%
J_{C}} - \mathbf{J_{L}}]  \tag{AI.3}
\end{equation}

Using the expression 4.15, we have the solution of the wave equation AI.3.
\begin{equation}
\mathbf{E_{C}}(R) - \mathbf{E_{L}}(R) = 4\pi (1-v^{2}/c^{2}) \int \frac{%
\nabla_{r} [\rho_{C}(r) - \rho_{L}(r)] } {\left |\mathbf{R}-\mathbf{r} \right|}
d\mathbf{r}  \tag{AI.4}
\end{equation}
which coincides to the Eq. 4.16.

Now we prove that the rhs of Eq. AI.4 asymptotically tends to zero. Because
for two arbitrary functions it is fulfilled relation
\begin{equation*}
\int F(\mathbf{R}- \mathbf{r})\nabla_{r} f(\mathbf{r})d\mathbf{r}
=\nabla_{R}\int F(\mathbf{R}-\mathbf{r}) f(\mathbf{r})d\mathbf{r}
\end{equation*}
we have for Eq. AI.4
\begin{equation}
\mathbf{E_{C}}(R) - \mathbf{E_{L}}(R) = 4\pi (1-v^{2}/c^{2}) \nabla_{R}\int
\frac{[\rho_{C}(r) - \rho_{L}(r)]} {\left |\mathbf{R}-\mathbf{r} \right|}
d\mathbf{r}  \tag{AI.5}
\end{equation}

Using the Eqs. 4.8 and 4.10 of \cite{JDJ}, we obtain for Eq. AI.5
\begin{equation}
\mathbf{E}_{C}(\mathbf{R},t)-\mathbf{E}_{L}(\mathbf{R},t)=4\pi 
\left( 1-\frac{v_{2}}{c_{2}}\right) \nabla _{R}\left[\frac{q_{C}}{R}
+\frac{(\mathbf{p}_{C}\mathbf{R})}{R^{3}}-\frac{q_{L}}{R}-
\frac{(\mathbf{p}_{L}\mathbf{R})}{R^{3}}+O(1/R^{3})\right] \tag{AI.6}
\end{equation}%
where $q_{C}$ and $q_{L}$, $\mathbf{p}_{C}$ and $\mathbf{p}_{L}$ are the
charges and electric dipole moments in both gauges. Since the charges are
identical in both gauges and absolute value of dipole moment of the charge
cannot be greater $aq$ , we have for Eq. AI.6
\begin{equation*}
\left| \mathbf{E}_{C}(\mathbf{R},t)-\mathbf{E}_{L}(\mathbf{R},t)\right|
<4\pi\left( 1-\frac{v_{2}}{c_{2}}\right) \frac{aq}{R^{3}}
\end{equation*}%
i.e. this term rapidly tends to zero from distances some times greater the
classical radius of the charge.

\begin{center}
\textbf{Appendix II. Obtaining of the shape of the elementary charge in the
Coulomb gauge.}
\end{center}

Here we analyze the following statement: if the moving charge acquires a
shape of contracted ellipsoid in the Lorentz gauge, will we observe the same
effect for the moving charge in the Coulomb gauge. It seems it must be so
because any physical quantity must transform according to the Lorentz
transformations while going from one inertial frame to the other one. But
without possibility to verify experimentally how the shape of the charge
actually changes, in the given gauge, the above statement can be treated
only as assumption.

However, a problem is just in this experimental verification since we are
not able to reconstruct the shape of the elementary charge directly from the
experimental data, i.e. from measured EM fields.

\begin{itemize}
\item For the moving charge, the lines of the E field are not normal to the
surface, which these lines outcome from (the example of such a configuration
of the lines of the E field and the moving charge is given in the Fig. 26.4
of \cite{Fey}), so we cannot use geometric methods

\item Formally the shape of the uniformly moving charge may be determined as
a solution of the integral equation for the electric and magnetic fields,
where the function r is a source for the Green function, but because we
cannot fix the radius of the elementary classical charge, the problem cannot
have unambiguous solution.
\end{itemize}

It follows from the pp. 1 and 2, that the only way to determine the shape of
the elementary charge is after equipotential surfaces of $\varphi$, i.e. the
way used by Lorentz. One can object that because the EM potentials are
treated, within the classical electrodynamics, as some abstract but not
physical quantities, unambiguous determination of the shape of the charge
via the scalar potential is impossible. However, in the above problem, just
the properties of the EM potentials are under investigation, therefore, for
us it is not so important what is an origin of the potentials. But what is
important is the fact that $\varphi$ and $\mathbf{A}$ are unambiguously
defined after the EM fields and the condition on a type of the gauge.
Therefore, the shape of the charge will be determined unambiguously too
because there is no ambiguity in the gauge condition.

Now we show by \textit{reductio ad absurdum} that the uniformly moving
elementary charge cannot have, in the Coulomb gauge, a shape of contracted
ellipsoid.\newline
Thus, in the frame with the charge at rest, the Poisson equation for the
elementary charge is (in Gauss units)\newline
\begin{equation}
\left( \frac{\partial ^{2}}{\partial (x^{\prime })^{2}}+\frac{\partial ^{2}}{%
\partial (y^{\prime })^{2}}+\frac{\partial ^{2}}{\partial (z^{\prime })^{2}}%
\right) \varphi ^{\prime }=-4\pi \rho (x^{\prime },y^{\prime },z^{\prime })
\tag{AII.1}
\end{equation}

and its solution is
\begin{equation}
\varphi^{\prime}(\mathbf{R}^{\prime}) = \int \frac{\rho
(x^{\prime},y^{\prime},z^{\prime})} {\left| \mathbf{R}^{\prime}-\mathbf{r}^
{\prime}\right|}dx^{\prime}dy^{\prime}dz^{\prime}  \tag{AII.2}
\end{equation}

We don't know what is the shape of the uniformly moving elementary charge
but we exactly know the shape of the charge while it is at rest. Due to
rotational symmetry, its shape \textit{must be spherical}.\newline
For the 'point-like' charge, i.e. for $R^{\prime}>> r^{\prime}$ the Eq. AII.2
reduces to
\begin{equation}
\varphi^{\prime}(\mathbf R^{\prime}) = \frac{q}{R^{\prime}}  \tag{AII.3}
\end{equation}

Now we go to the second frame where the charge moves uniformly. According to
the Lorentz transformations, the function $\rho$ must transform and the
shape of the elementary charge becomes elliptic.

But when we apply the Lorentz transformation to the physical quantities,
even they are treated as auxiliary ones, in some frame, we must transform
\textit{all} quantities of this frame. So in the Eqs. AII.1, AII.2 and AII.3,
we must transform not only the potentials and the charge densities but the
coordinates too.\newline
As a result, we have
\begin{equation*}
\varphi ^{\prime }(R)=\frac{q}{\sqrt{(x-vt)^{2}+(1-v^{2}/c^{2})(y^{2}+z^{2})}}
\end{equation*}%
But this solution does not coincide to well known solution for the scalar
given in \cite{JDJ} (Eq. 6.45).
\begin{equation*}
\varphi (\mathbf{R})=\int \frac{\rho (x,y,z,t)}{\left|
\mathbf{R}-\mathbf{r}\right| }dxdydz
\end{equation*}%
that is correct for any law of motion of the charge. So our suggestion about
the Lorentz contraction of moving elementary charge leads to incorrect
expression for the $\varphi $ potentials and, therefore, any analysis of the
equivalence of the Coulomb and Lorentz gauges, which is presented in Refs.
2, 3 and 4 too, loses its significance.\newline
Thus, we are not able to conclude that the shape of the moving charge in the
Coulomb gauge is elliptic.

At the end, we prove that in the Coulomb gauge, the uniformly moving charge
has spherical shape.

It is easily to see that the equation for the scalar potential in the
Coulomb gauge obeys the 'Lorentz transformations' in the limiting case when
the speed of the scalar EM interaction tends to infinity $\left(
c\rightarrow \infty \right) $ so the Lorentz transformations reduce to
\begin{equation*}
x^{\prime}= \left( x-vt \right)/\sqrt{1-v^{2}/c^{2}} \rightarrow x-vt
\end{equation*}
Then the formula for transformation of the charge density becomes
\begin{equation*}
\rho^{\prime}= \rho /\sqrt{1-v^{2}/c^{2}} \rightarrow \rho
\end{equation*}
and the spherical shape of the uniformly moving charge remains to be
spherical too

\bibliographystyle{plain}

\end{document}